\documentstyle[prd,aps,floats]{revtex}

\newcommand{\be}{\begin{equation}}
\newcommand{\ee}{\end{equation}}
\newcommand{\ben}{\begin{eqnarray}}
\newcommand{\een}{\end{eqnarray}}
\newcommand{\bc}{\begin{center}}
\newcommand{\ec}{\end{center}}

\begin{document}

\draft

\input epsf
\renewcommand{\topfraction}{1.0}

\preprint{gr-qc/9802054}

\title{\bf Gravitational evolution and stability of boson halos}

\author{
Jayashree Balakrishna$^{1}$\footnote{Electronic address: jab@howdy.wustl.edu}
and
Franz E.~Schunck$^{2}$\footnote{Electronic address: fs@astr.cpes.susx.ac.uk}}

\address{$^{1}$McDonnell Center for the Space Sciences,
Washington University, St. Louis, MO 63130}
\address{$^{2}$Astronomy Centre, School of Chemistry, Physics and
Environmental Science, University of Sussex, Falmer, Brighton BN1 9QJ,
United Kingdom}


\maketitle

\begin{abstract}
We investigate the gravitational evolution of dark matter halos made up of
a massless bosonic field. The coupled Einstein-Klein-Gordon equations
are solved numerically, showing that such a {\em boson halo} is stable
and can be formed under a large
class of initial conditions. We also present an analytical proof that such
objects are stable in the Newtonian limit. 
In the context of boson stars made of massive scalar fields, we introduce
new solutions with
an oscillatory scalar field, similar to boson halos.
We find that these solutions are unstable.
\end{abstract}



\section {\bf Introduction}

Boson stars are self-gravitating configurations which are described
mathematically by the Einstein-Klein-Gordon equations \cite{rep}.
In order to find such
globally regular solutions it is necessary to have a conserved particle
number which is fulfilled if one uses a complex scalar field and a
potential with global $U(1)$ symmetry. Investigations so far have shown
that stable and unstable configurations exist so that study to a greater
extent makes sense \cite{kus1,kus2,sei}.
A boson star can have values of the mass over large orders
of magnitudes; it depends on the mass of the scalar field and
possible self-interactions \cite{col}. A recent paper described the effects if
the scalar field would interact only gravitationally so that
the star is transparently similar as a dark matter halo of a galaxy \cite{lid}.
Baryonic matter could then move within the gravitational potential of the
star and if perhaps a accretion disk can be built the radiation
would suffer a gravitational redshift. Redshift of an iron-K line
is actually observed within the center of a Seyfert galaxy \cite{iwa}.

Boson star solutions are characterized by an exponential decrease of
the scalar field for which a mass term in the potential is responsible.
A zero temperature solution is usually considered
in which all scalar particles are within the same ground state forming a
Bose-Einstein condensate. The size of this general-relativistic object
is given by the Compton wavelength. If one `switches off' the mass of
the scalar field, the Compton wavelength is infinite. Then the massless scalar
particles are within a coherent state, a kind of `boson star',
which in principle has an infinite range. Investigations of these solutions
\cite{sch1,sch2,sch3,sch4} which we call now {\em boson halos},
have shown that the mass of these objects increases linearly, a behavior
suggested by HI observations for spiral and dwarf galaxies.
It was shown that the rotation curves for galaxies could be very well
fitted by this kind of matter. In this paper, we present an analytical
proof showing the stability of the Newtonian solutions.

In this paper we also examine numerically how these objects could have formed
and show that they
are stable. For this purpose, we use the process of gravitational cooling
as described in \cite{sei2}. We add a small radial perturbation to a
solution and show that the evolved solution does not disperse or form a
black hole, hence it is
stable. Furthermore, we present here for the first time solutions with
an oscillating scalar field for the massive case; we find that they are
unstable.

In section II we present the mathematical foundations of the problem of
bosonic objects made from massless scalar fields. We set
up equilibrium equations and approximate solutions. We then set up the
evolution equations which we use in our code. In section III we discuss
our numerical studies on the formation of scalar halos followed in section
IV by an analytical proof of stability of these objects. In section V we
discuss our numerical studies of stability of bosonic halos. In section VI
we discuss new oscillatory (in space) solutions for massive scalar fields.
Finally we have a discussion of our results and their implications in
section VII.

\section{Mathematical Foundations of the Problem}

In this Section we start with the Lagrangian for the system and set up
the equilibrium equations and their solutions as well as the evolution
equations with which we evolve the equilibrium solutions.
\subsection {\bf Einstein-scalar-field equations}

The Lagrange density of a complex self-gravitating scalar
field reads
\be
{\cal L} = \frac {1}{2} \sqrt{\mid g \mid} \left [
  \frac {1}{\kappa } R + g^{\mu \nu } (\partial_\mu \Phi^\ast)
  (\partial_\nu \Phi) - U(|\Phi|^2) \right ] \; , \label{lagr}
\ee
where $R$ is the curvature scalar, $\kappa = 8\pi G$, $G$ the
gravitation constant ($\hbar=c=1$),
$g$ the determinant of the metric $g_{\mu \nu }$,
$\Phi $ the complex scalar field, and $U$ the potential depending on
$|\Phi|^2$ so that a global $U(1)$ symmetry is conserved.
Then we find the coupled system
\ben
R_{\mu \nu } - \frac{1}{2} g_{\mu \nu } R  & = &
                  - \kappa T_{\mu \nu } (\Phi ) \; , \\
\Box \Phi + \frac{\partial U}{\partial \Phi^\ast} & = & 0 \; ,
\een
where
\be
T_{\mu \nu } = (\partial_\mu \Phi^\ast ) (\partial_\nu \Phi )
 - \frac{1}{2} g_{\mu \nu }
 [ g^{\sigma \kappa } (\partial_\sigma \Phi^\ast )
         (\partial_\kappa \Phi ) - U(|\Phi|^2) ]
\ee
is the energy-momentum tensor and
\be
\Box = \partial_\mu
\Bigl [ \sqrt{\mid g \mid } g^{\mu \nu } \partial_\nu \Bigr ]/ 
\sqrt{\mid g \mid }
\ee
the generally covariant d'Alembertian.

For spherically symmetric solutions we use the following line element
\be
ds^2 = N^2({\bf r},{\bf t}) d{\bf t}^2 - g^2({\bf r},{\bf t}) d{\bf r}^2
  - {\bf r^2} ( d\vartheta^2 + \sin^2\vartheta \, d\varphi^2) \label{metric}
\ee
and for the scalar field the ansatz
\be
\Phi ({\bf r},{\bf t}) = P({\bf r},{\bf t}) e^{-i \omega {\bf t}} \; , \\
\ee
where $\omega $ is the frequency.

\subsection{Equilibrium Equations}

The non-vanishing components of the energy-momentum tensor are
\ben
T_0{}^0 = \rho & = & \frac{1}{2} \left [ \frac{\omega^2  P^2}{N^2}
   + \frac{P'^2}{g^2} + U \right ] \; , \\
- T_1{}^1 = p_{\bf r} & = &
 \frac{1}{2} \left [ \frac{\omega^2  P^2}{N^2}
   + \frac{P'^2}{g^2} - U \right ] \; , \\
- T_2{}^2 = - T_3{}^3 = p_\bot & = &
 \frac{1}{2} \left [ \frac{\omega^2  P^2} {N^2}
   - \frac{P'^2}{g^2} - U \right ] \; ,
\een
where $'=d/d{\bf r}$.

The equilibrium configurations (those for which the metrics are static)
for a system of massless scalar fields are derived from the following
equations
\begin{equation}
\sigma' = \chi \; , \label{chii}
\end{equation}
\begin{equation}
\chi'= -\left[\frac{1}{r}+\frac{g^2}{r}\right]\chi -
\frac{\sigma g^2}{N^2} \; ,
\end{equation}
\begin{equation}
g'=\frac{1}{2} \left[\frac{g}{r}-\frac{g^3}{r}+\frac{\sigma^2 r g^3}{N^2}
+ r g \chi^2 \right] \; ,
\end{equation}
\begin{equation}
N' =\frac{1}{2}\left[-\frac{N}{r}+\frac{N g^2}{r}+\frac{r g^2 \sigma^2}{N}
+ r N \chi^2 \right] \; . \label{chif}
\end{equation}
Here we use the dimensionless variables $r=\omega {\bf r}$,
$t=\omega {\bf t}$ and $\sigma = \sqrt{4\pi G\, } \phi$.
Regularity at the center implies $g(r=0)=1$.
The equilibrium configurations are characterized by saddle points in the
density $\rho$, which might
have some significance in stability issues. These systems are characterized
by a two parameter family
of solutions. Firstly, we can vary $\sigma(0)$, the central density,
and secondly, for
each $\sigma(0)$ we can vary $N(0)$. Although we do not have asymptotic
flatness (the mass
$M$ rises with $r$) the mass profile for a given value of $r$ has a very
interesting feature. 
Figure $1a$  shows the mass as a function of central density for a given
value of $r=R$ and
$N(0)$. The mass throughout the paper is calculated from the Schwarzschild
metric
\begin{equation}
M(r)=\frac{r}{2} \left ( 1 - \frac{1}{g^2} \right ) \; ,
\end{equation}
where $g^2$ is the radial metric.

The mass profile is very similar to that of the massive scalar field case
although there the mass is not subject to these variations with $r$.
This is also reminiscent of neutron star
profiles. The increase in mass with increase in central density is
followed by a decrease in mass
with further increase in central density. 
For the same value of $N(0)$ and larger values of $r$ the mass increases
but the profile remains the same. 
Figure $1b$ shows mass versus radius for different central
densities and a given value of $N(0)$. The profile is independent of the
value of the radius for large values of $r$.

The Noether theorem associates with each symmetry a locally conserved 
``charge''. The Lagrangian density (\ref{lagr}) is invariant under a 
global phase transformation
$\Phi \rightarrow \Phi e^{-i\vartheta }$. The 
{\em local conservation law} of the associated Noether current 
density reads
\be
\partial_\mu j^\mu =0\; , \qquad
j^\mu = \frac {i}{2} \sqrt{\mid g\mid }\; g^{\mu \nu }
 [\Phi^\ast \partial_\nu \Phi -\Phi \partial_\nu \Phi^\ast ] \; . \label{Noet}
\ee
If one
integrates the time component $j^0$ over the whole space we find the
{\em particle number}
\be
N_p = 4\pi \omega \int\limits_0^\infty \; \frac{g}{N} r^2 P^2 \, dr
\label{particle}  \; .
\ee

\subsection {Approximate solution}

The Newtonian and the general relativistic solutions of system
(\ref{chii})-(\ref{chif}) for $U=0$ were recently discussed
\cite{sch1,sch2,sch3,sch4}.
The Newtonian ones (initial value $\sigma (0)$ smaller than about $10^{-2}$)
can be used to describe dark matter halos of spiral and
dwarf galaxies. The general relativistic ones
(initial value $\sigma (0)>10^{-2}$) reveal very high
redshift values. As was shown in \cite{sch1,sch2,sch3,sch4},
for the Newtonian solutions
there exists an approximately analytic formula for the scalar field
($N=g=1$)
\be
\sigma (r) = A \frac{\sin (r)}{r} \; ,
\label{sigma}
\ee
where $A$ is a constant. The energy density reads
\be
\rho (r) =  \frac{A^2}{r^2} \left [ 1 - \frac{\sin (2 r)}{r}
+ \frac{\sin^2 (r)}{r^2} \right ]  \label{rho}
\ee
and the mass defined as usual as
$M(r) = \int_0^r \rho (\zeta ) \zeta^2 d\zeta $ yields
\be
M(r) =  A^2 \left [ r + \frac{\cos (2 r) - 1}{2 r} \right ]
 \label{mass} \; .
\ee
The energy density shows a decreasing behavior with saddle points in between
(cf.~Fig.~3a).
The Newtonian solution for the particle number is
\be
N_p(r) =  \frac{A^2}{4} \biggl [ 2 r - \sin (2 r) \biggr ]
 \label{part2} \; ,
\ee
the quantitative behavior of which is also revealed by numerical calculation.

\subsection{Evolution Equations}

For accurately numerical evolution the following set of variables are chosen
\cite{sei}
\begin{equation}
\psi_1 \equiv r\sigma_1 \; ,\quad
\psi_2 \equiv r\sigma_2 \; ,\quad
\pi_1 \equiv \frac{1}{\alpha}\frac{\partial\psi_1}{\partial t} \; ,\quad
\pi_2 \equiv \frac{1}{\alpha}\frac{\partial\psi_2}{\partial t} \; ,
\end{equation}
where
\begin{equation}
\alpha \equiv \frac{N}{g} \; ,
\end{equation}
and the subscripts on $\psi_i$ denote the real and imaginary parts of the
scalar field multiplied by $r$.

In terms of these variables and the dimensionless ones in the previous
Section the evolution equations are as follows: The radial metric function
$g$ evolves according to
\begin{equation}
{\dot g}=N(\pi_1\sigma_1'+\pi_2\sigma_2') \; .
\end{equation}
The polar slicing equation, which is integrated on each time slice,
is given by
\begin{equation}
N'=\frac{N}{2}\left[\frac{g^2-1}{r}+r\left[(\sigma_1')^2+(\sigma_2')^2
\right]+\frac{\pi_1^2+\pi_2^2}{r}
\right] \; .
\end{equation}
The Klein-Gordon equation for the scalar field can be written as
\begin{equation}
{\dot\pi_i}=\alpha'\psi_i'+\alpha\psi_i''-
 \psi_i\left[gN+\frac{\alpha'}{r}\right]
,\quad i=1,2 \; ,
\end{equation}
\begin{equation}
{\dot\psi}_i=\alpha\pi_i,\quad i=1,2 \; .
\end{equation}
The Hamiltonian constraint equation is given by
\begin{equation}
\frac{2g'}{rg^3}+\frac{g^2-1}{r^2g^2}-\frac{\pi_1^2+\pi_2^2}{r^2g^2}-
\frac{\sigma_1'^2+\sigma_2'^2}{g^2}
=0 \; .
\end{equation}

After we introduce a perturbation in the field or when we have
an arbitrary initial field configuration we reintegrate on
the initial time slice to get new metric components \cite{sei}.

\section{Formation of Scalar Objects}

In this Section we discuss the possibility of formation of massless
scalar field configurations
from non-equilibrium data. Earlier studies both numerical and analytical
have confirmed that
self-gravitating objects with massless scalar fields cannot be compact
\cite{sei2,chr}. Thus taking a Gaussian type
localized distribution of scalar matter and evolving the system using the
evolution equations
described above, results in dissipation of the scalar matter without forming
any self-gravitating
object. Even sinusoidal functions that were damped by exponential decays met
with the same fate. 
These were functions like $\exp(-r)\sin(r)/r$ and $\exp(-r)\cos(r)$.
In Figure 2, a plot of the
density versus the radius for different times is shown for
$\sigma = 0.001 \cos (r) \exp(-r)$.
The star dissipates very quickly. 

The initial configurations for the scalar field that yield stable
configurations were those
characterized by a $1/r$ times a sinusoidal dependence at
large $r$ as well as a saddle point structure of the energy density
$\rho$. These were typically functions like
\be
\frac{\cos(r)}{1+r} \; , \quad
\cos(r) \left [ 1-\exp \left (-\frac{1}{r} \right ) \right ] \; ,
\quad \mbox{and}
\ee
\be
\sin(r) \left ( 1+ \frac{1}{r} \right ) \log \left [1+ \frac{1}{1+r} \right ]
\; .
\ee
In Figure 3 we show the density and radial metric evolutions
for a field of the form $0.003\cos(r) (1-\exp(-1/r))$. This settles
into a self-gravitating
object after some time. Figure $3a$ shows the density as a function of
$r$. The central density $\rho(0)$ increases during the evolution
from its initial value at $t=0$. Figure $3b$ reveals the radial metric as it
evolves in time as a
function of radius. This too displays the settling to a stable configuration.
In Figure $3c$ we show the mass loss for the system as it finds itself
a configuration. The mass at fixed radial values for these times is
shown in Table 1.
The amount of radiation for this system is relatively small and decreases in
time.

On the other hand, functions like $1/(r+1)$ , $\cos (r)/{(1+r)}^2$ and
$\sin (r^{1.01})/r^{1.01})$ failed
to settle to a bound state and just dissipated away. These functions did
not have saddle points in $\rho$. 

Functions like $\sigma=\sin (r)/\log (1+r)$  for which $\rho$ 
has saddle points (it also has a maximum near the origin), did not
dissipate but failed to settle in a
long numerical evolution. This might mean they would form the halo
structures in a very long time 
for which it would not be computationally feasible to evolve the system.
The same was observed for $\sin (r)/(r \log (1+r))$ or a numerical
six node boson star configuration ($\omega <m$).

All the above configurations were fairly Newtonian for which the exact
solution was close to $\sin (r)/r$.

\section{Analytical stability proof}

We investigate here the stability of the Newtonian solutions against
small radial perturbations; for stability investigations concerned with
the boson star model and using a perturbative method,
see Ref.~\cite{LP89,J89,GW89}. Therefore, we can neglect the
perturbations for the spacetime and make the following ansatz for the
linear scalar field perturbations $\delta P$:
\be
\Phi (r,t):=P(r) e^{-i\omega t} + \delta P(r) e^{i k_n t} \; ,
\ee
where $k_n$ are the frequencies of the normal modes. The pulsation equation,
an eigenvalue equation, is then:
\be
\delta P''+2 \delta P'/x+k_n^2 \delta P = 0 \; .  \label{deltap}
\ee
Together with the boundary conditions
\be
\delta P'(0)=\delta P (R)=0
\ee
for regularity of $\delta P$ at the origin and at the radius $R$ of the
boson halo, this is a Sturm-Liouville eigenvalue problem.
As it is well-known from mathematical theory, we have a series of
real eigenvalues with a minimal one:
\be
k_1^2 < k_2^2 < \ldots  \; .
\ee
Eigenfunctions of this particular differential equation (\ref{deltap})
exist only if the eigenvalues $k_n^2$ are positive.
This means that all modes are stable. The eigenfunctions are
\be
\delta P = \frac{\sin(k_n x)}{x} \; ,
\ee
where the eigenvalues are
\be
k_n = n \frac{\pi }{R} \; ;
\ee
$n$ is a natural number and $R$ the radius of the solution.

\section{Numerical Stability Issues}

In order to study the stability of these systems numerically
we perturbed exact configurations and observed whether
the system settled to new configurations. In some sense taking a function like
$A \sin(r)/r$ which is quite close to the exact solution for a Newtonian system
or $A \cos(r)/(1+r)$ can be regarded as a perturbation on the exact
solution and the eventual
settling to a new configuration can be regarded as indicative of stability. 
The settling of $0.001\cos(r)/(r+1)$ to a stable configuration is shown
in Figure 4.
The density at the center increases in this case from what it starts at.
The radial metric evolution case is shown in Figure 4b.
On the other hand when $A=1$ the system is non-Newtonian
and in a long evolution failed to settle down although it did not disperse. 
This is not surprising since the $\sin (r)/r$ form is close to the solution
only in the Newtonian case.

In Figure 5 we show a perturbed Newtonian configuration as it settles to
a new Newtonian configuration.
The perturbation that is used in this case mimics an annihilation of
particles. A Gaussian bump
of field is removed from a part of the star near the origin.
Figure 5a is a plot of the unperturbed versus the perturbed density at
$t=0$. The perturbed configuration
is evolved and settles to a new configuration. The scalar radiation moves
out as shown in Figure 5b.
In Figure 5c the density profile is shown after the system settles down.
The system is very
clearly in a new stable configuration. In Figure 5d the mass is plotted
as a function of radius
for various times. Again one can see that the mass loss is decreasing by
the end of the run showing
that the system is settling to a new configuration. The mass as a
function of time is presented in Table 2 for different radii.

We have so far been successful only in our Newtonian evolutions.
The reason for this is that denser configurations
need much better resolution for evolution. This is coupled with
the difficulty that we still
need the boundary to be very far away so that the density has significantly
fallen off. We are working on improving our boundary conditions.
So far, we are using either an outgoing wave condition
or exact boundary conditions where the latter one simulates a vacuum
energy. Further investigation is needed before we
can decide whether the
non-Newtonian configurations are inherently unstable.

\section {Boson star oscillators}

A boson star consists of scalar massive particles, hence in the simplest
model one has a potential $U=m^2 |\Phi|^2$. Exponentially decreasing
solutions exist for special eigenvalues $\omega < m$,
so that the star has a finite mass.
In the case of $\omega > m$, oscillating scalar field solutions can
be found for all values of $\omega $. The energy density reveals
minima and maxima as opposed to the saddle point structure seen for
the massless case.
Figure 6a shows a comparison of the density profile for the two cases.

In figure 6b we show the mass profile and the particle number as functions
of central density. The binding energy is always positive making the
system unstable against a collective transformation in which it disperses
into free particles but the system is still stable against any one-at-a-time
removal of a particle \cite{harr}. The complete dispersion is a numerical
confirmation of the result by Zeldovich, cf.~\cite{zel}.

\section {\bf Discussion}

We have shown that massless complex scalar fields are able to settle down
to a stable configuration. It seems that the formation process needs
a special form for the energy density. The appearance of saddle
points within the decreasing density supports the `birth' of the boson halos.
In contrast, extremal values for the density cannot be compensated
and lead to the destruction of the initial configuration. This result
of our numerical investigation reveals that the formation of
boson halos could be difficult. But as our attempt by cutting off a part of a
Newtonian solution has shown, if there exists such a structure, then a
new configuration forms and settles down to a stable dark matter halo.
One can understand this as Bose-Einstein condensation where a small
initial configuration forms and particles in the surrounding falls into
the most favorable state given by the condensation. In this way, the
boson halo forms and grows up to the point where it is stopped by
the outer boundary, the vacumm energy density.
Our choice of the numerical boundary
was necessary so that it simulates a hydrostatic equilibrium with
this vacuum energy as it was assumed in \cite{sch2,sch3,sch4}.
Because the boundary can be placed at every radius the size of these
dark matter halo depends on the value of the cosmological constant.

During the formation process of the boson halo, massless scalar particles
change into the Bose-Einstein condensate and loose energy through
this process; only particles with at least this energy can participate.
We infer that the condensed state is the
energetically more preferable state in comparison with the free state.
The calculation of the binding energy, the comparison of mass of the
bounded with the free particle at infinity, is not advised due to
the arbitrariness of the energy of a massless particle at infinity.

Investigations with real scalar fields \cite{chr} or with different
boundary conditions \cite{sei} cannot lead to singularity
free solutions;
cf.~Liddle and Madsen in \cite{rep}.

The second class of solutions presented here is the oscillatory one for
a massive complex scalar field. Our result of a positive binding energy
confirms the expectations for systems where the energy of the field
is higher than the rest mass energy. Our numerical evolution also
reveals the instability by dispersion of the
unperturbed exact solution.

\acknowledgments
We would like to thank John D.~Barrow,
Andrew R.~Liddle, Eckehard W.~Mielke, Ed Seidel, Wai-Mo Suen,
and Pedro T.P.~Viana for helpful discussions and comments.
Research support of FES was provided by an European Union Marie Curie
TMR fellowship.
FES wishes to thank Peter Anninos and Wai-Mo Suen for their hospitality
during a stay at the University of Champaign-Urbana and at the Washington
University of St.~Louis.
Research of JB is supported in part by the McDonnell
Center for Space Sciences, the Institute of Mathematical Science of
the Chinese University of Hong Kong, the NSF Supercomputing
Meta Center (MCA935025) and National Science Foundation (Phy 96-00507).



\begin{figure}
\centering
\leavevmode\epsfysize=7.5cm \epsfbox{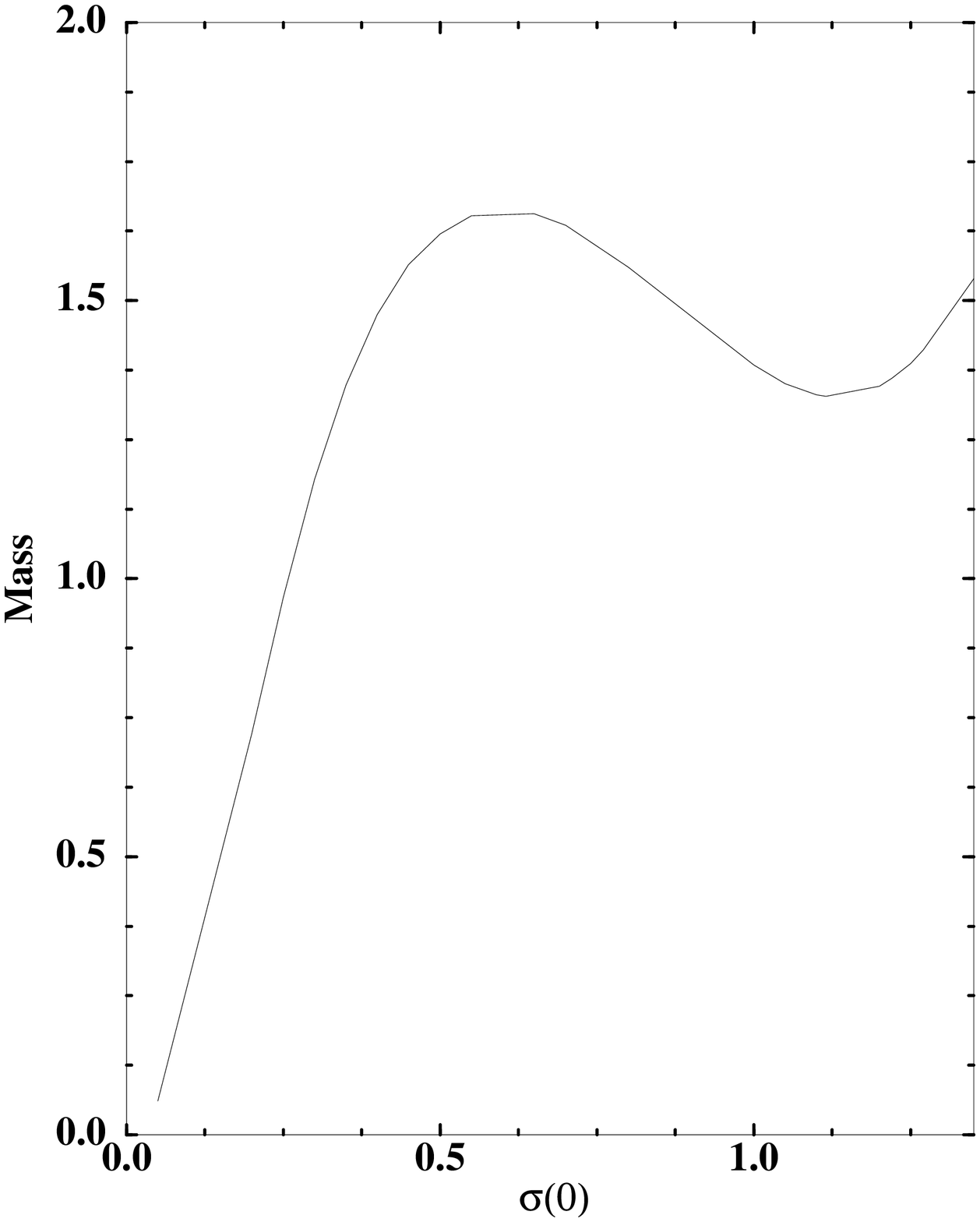}\hskip0.5cm
\leavevmode\epsfysize=7.5cm \epsfbox{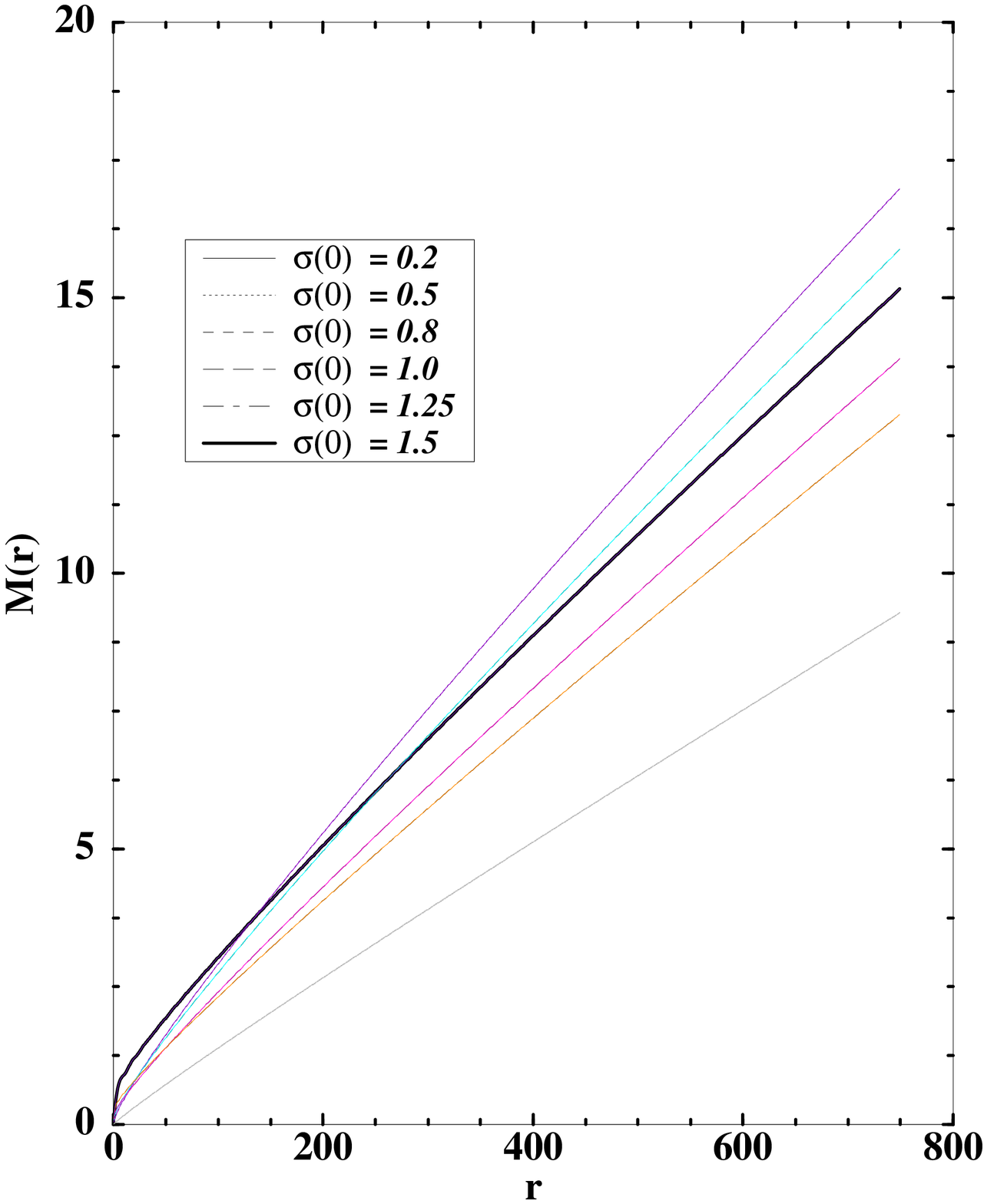}\hskip0.5cm
\leavevmode\epsfysize=7.5cm \epsfbox{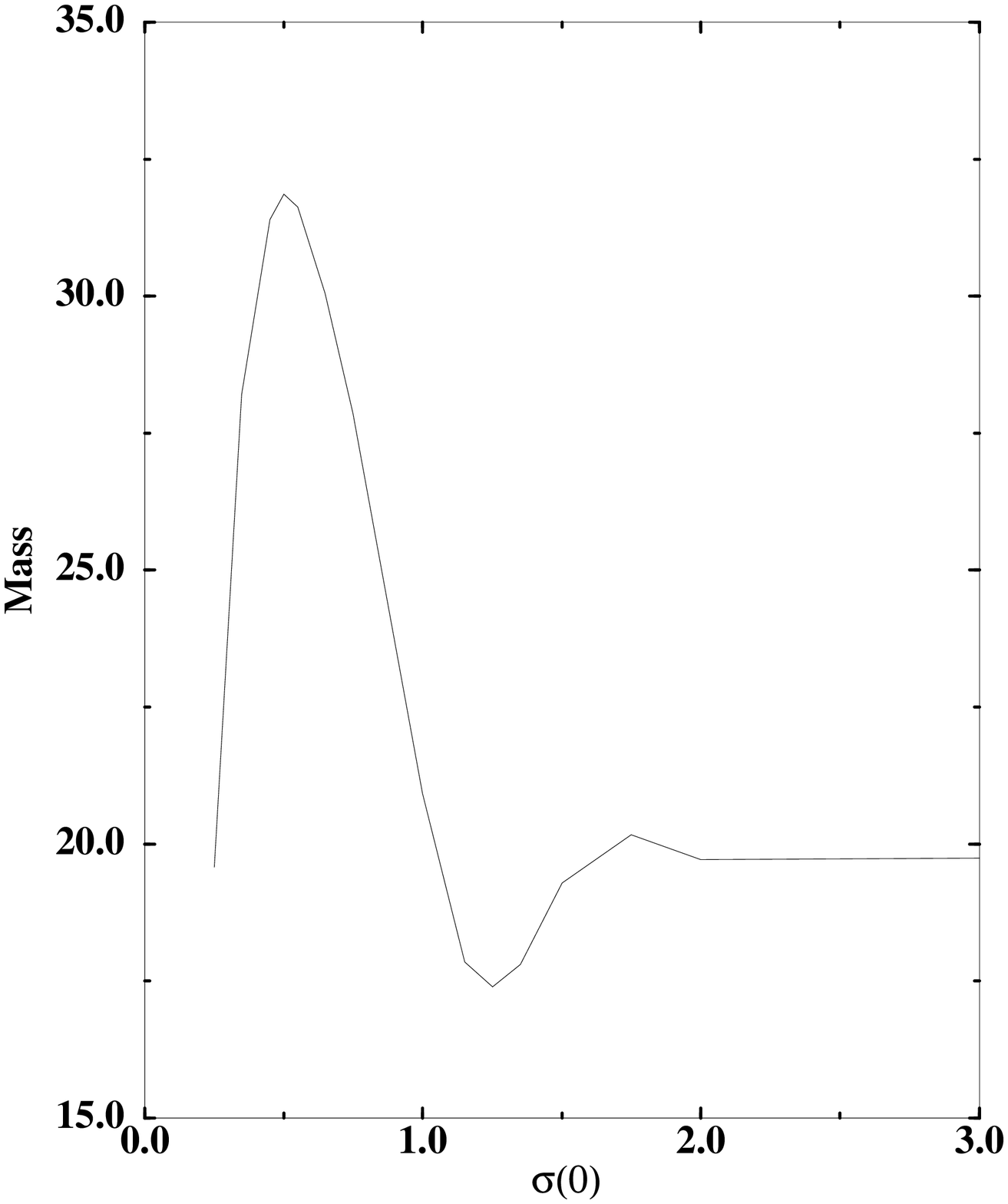}\\
\caption[]
{(a) Left: The mass profile of equilibrium configurations of massless scalar
fields, are interestingly
similar to the profiles of massive scalar fields and neutron stars.
The mass for a given radius
increases to a maximum with central density before it decreases with
further increase in
central density. However by calculating the particle number, we see
no division into stable and unstable
configurations at the peak. The particle number depends very sensitive
on the initial value of the lapse function $N$, so that a clear cusp
structure in the mass-particle-number diagram found for boson stars
cannot be derived here (cf.~\cite{kus1,sch1}).
(b) Middle: The profiles are largely independent of radius although the mass
itself increases with $r$.
(c) Right: If one cuts the boson halos at the same energy density, then the
mass profile reveals the same oscillating behavior as in case (a).
The energy density $\rho $ is measured in $\omega^2/\kappa $.}
\end{figure}

\begin{figure}
\centering
\leavevmode\epsfysize=7.5cm \epsfbox{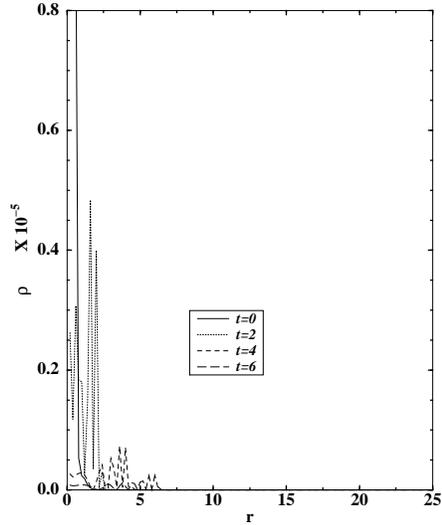}\\
\caption[]
{The inability of massless scalar field configurations to form
compact self-gravitating objects as
discussed by Christodoulou and others \cite{sei2,chr} is verified numerically.
An initial field configuration of the
form $\sigma = 0.001 \cos(r) \exp(-r)$ is seen dispersing in the plot.
The dispersion takes place very quickly.}
\end{figure}

\begin{figure}
\centering
\leavevmode\epsfysize=7.5cm \epsfbox{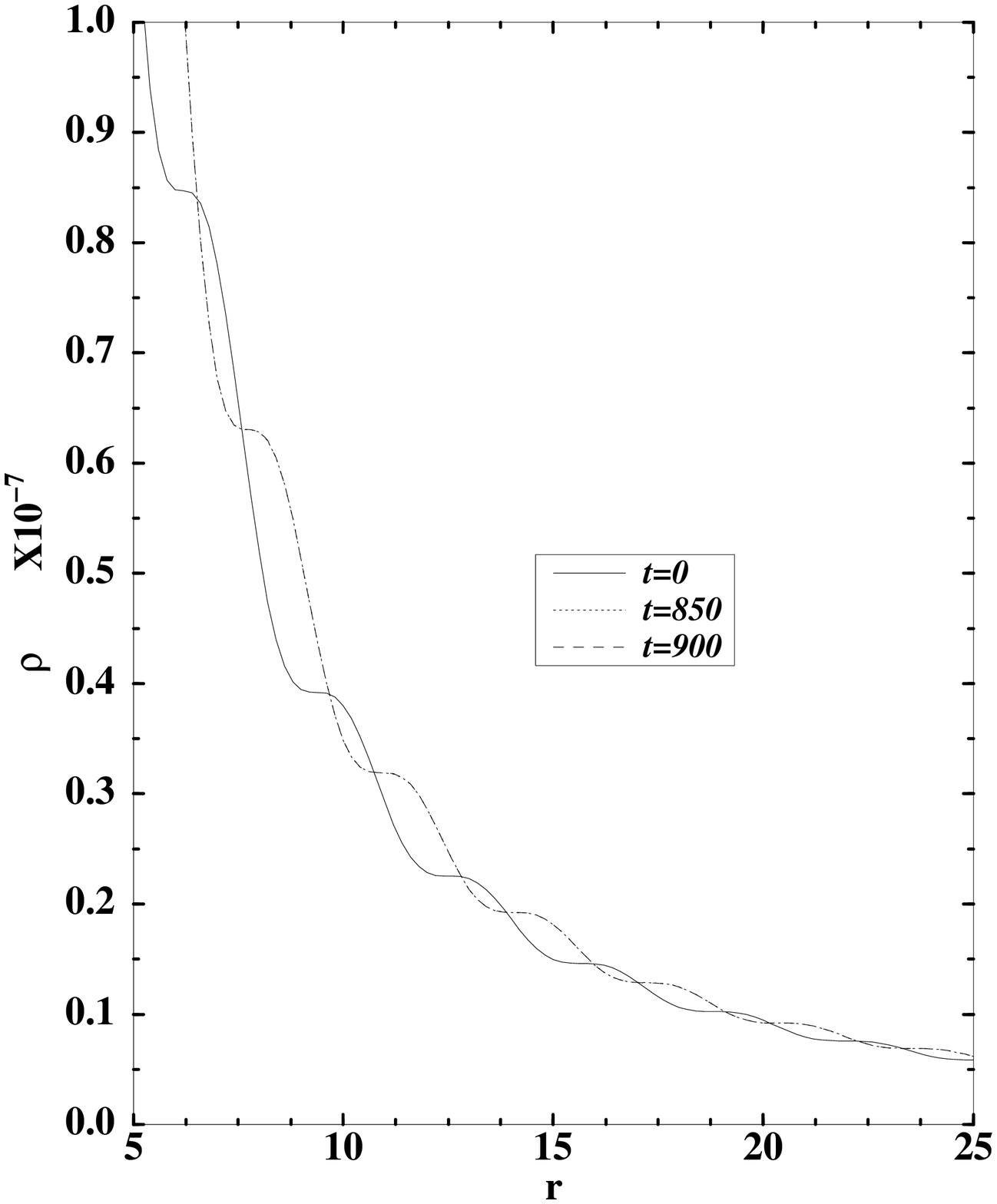}\hskip0.5cm
\leavevmode\epsfysize=7.5cm \epsfbox{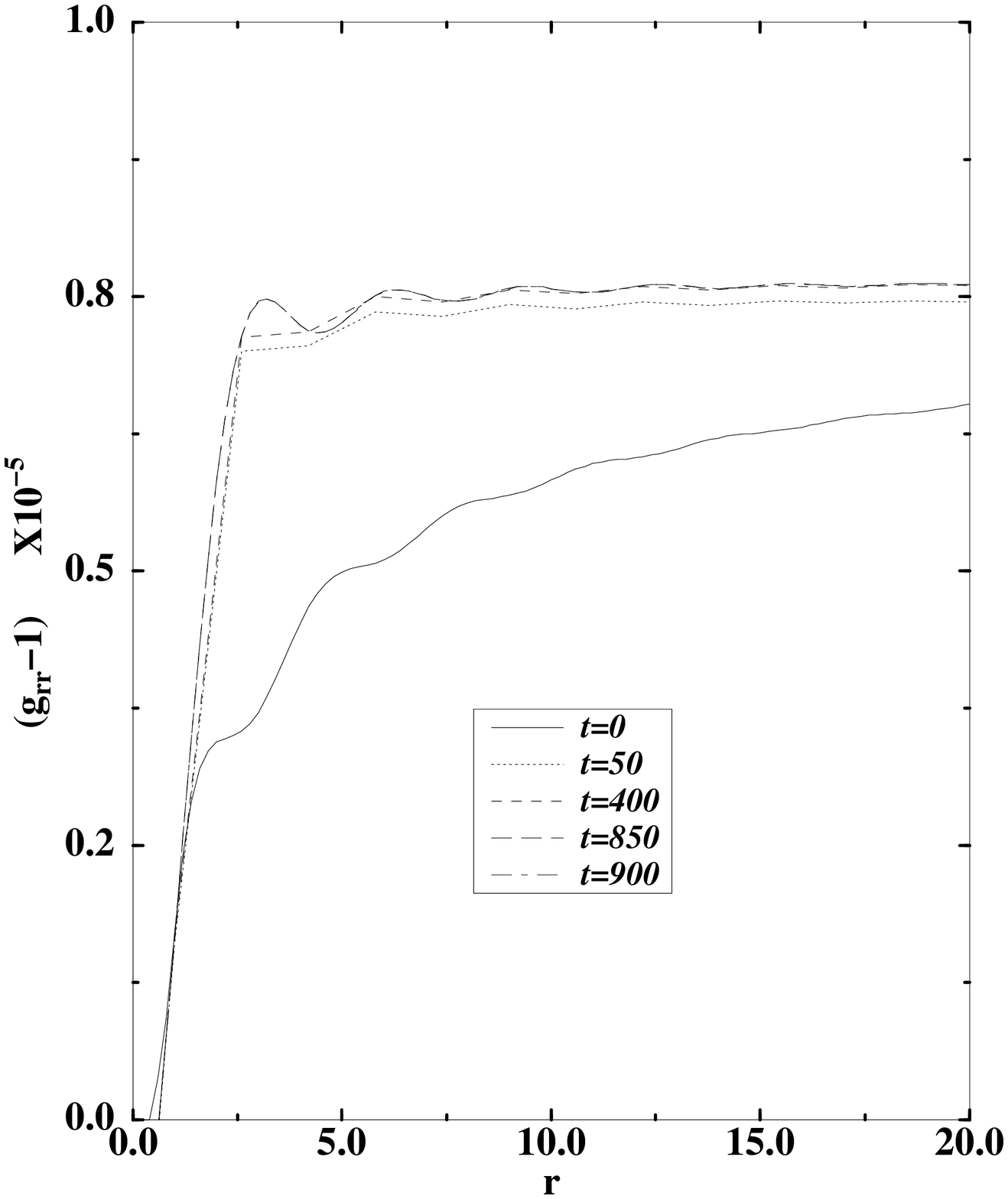}\hskip0.5cm
\leavevmode\epsfysize=7.5cm \epsfbox{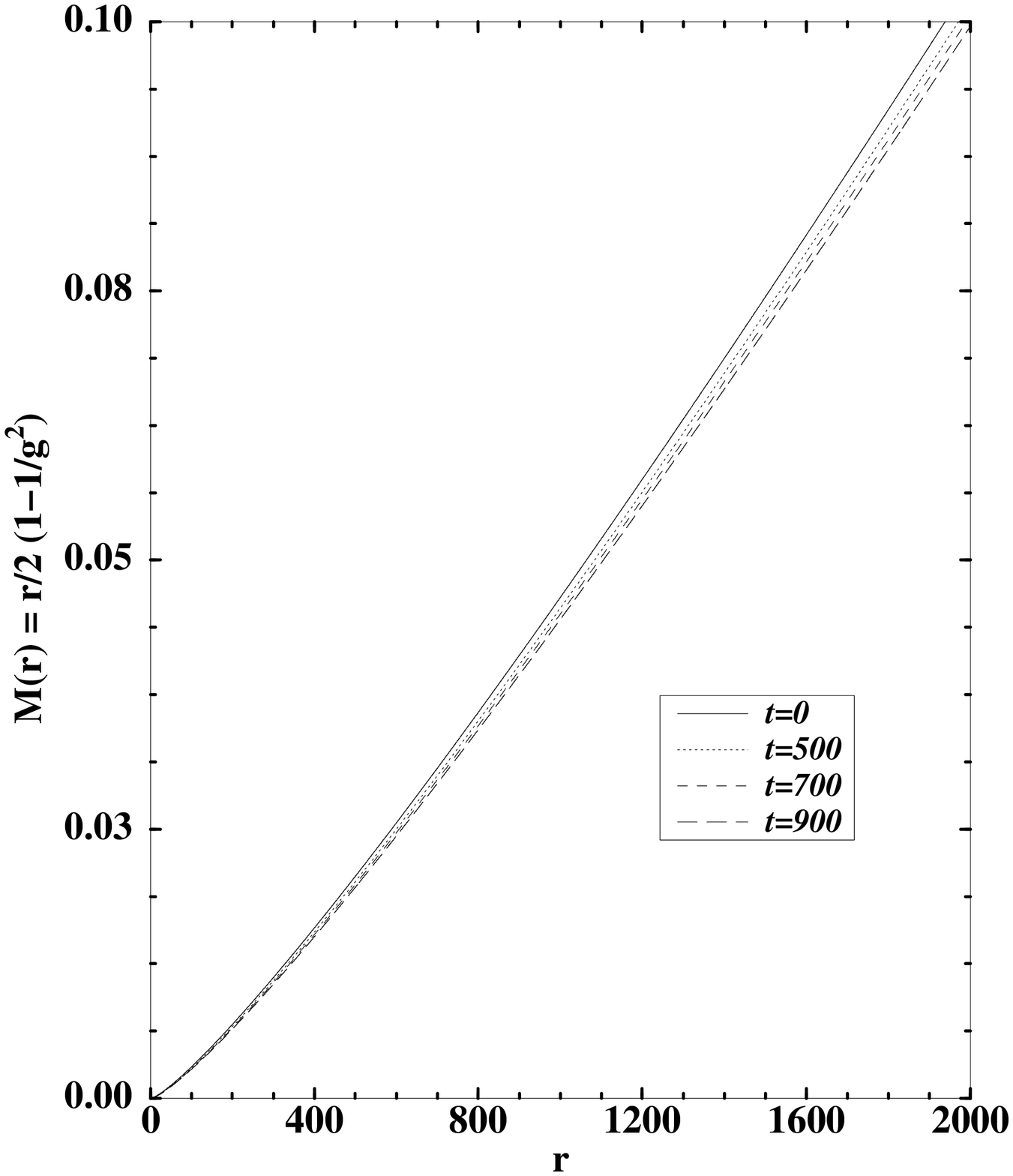}\\
\caption[]
{(a) Left: The formation of a self-gravitating massless scalar field object is
shown here for
an initial field configuration of the form $0.003 \cos (r) (1-\exp(-1/r))$.
The density in this
case increases from its initial value as the configuration settles down. 
(b) Middle: The radial metric for this configuration is shown evolving
to a stable final configuration.
(c) Right: The mass is plotted as a function of radius at different times.
As the configuration evolves it looses less and less mass as it settles down.}
\end{figure}

\begin{figure}
\centering
\leavevmode\epsfysize=7.5cm \epsfbox{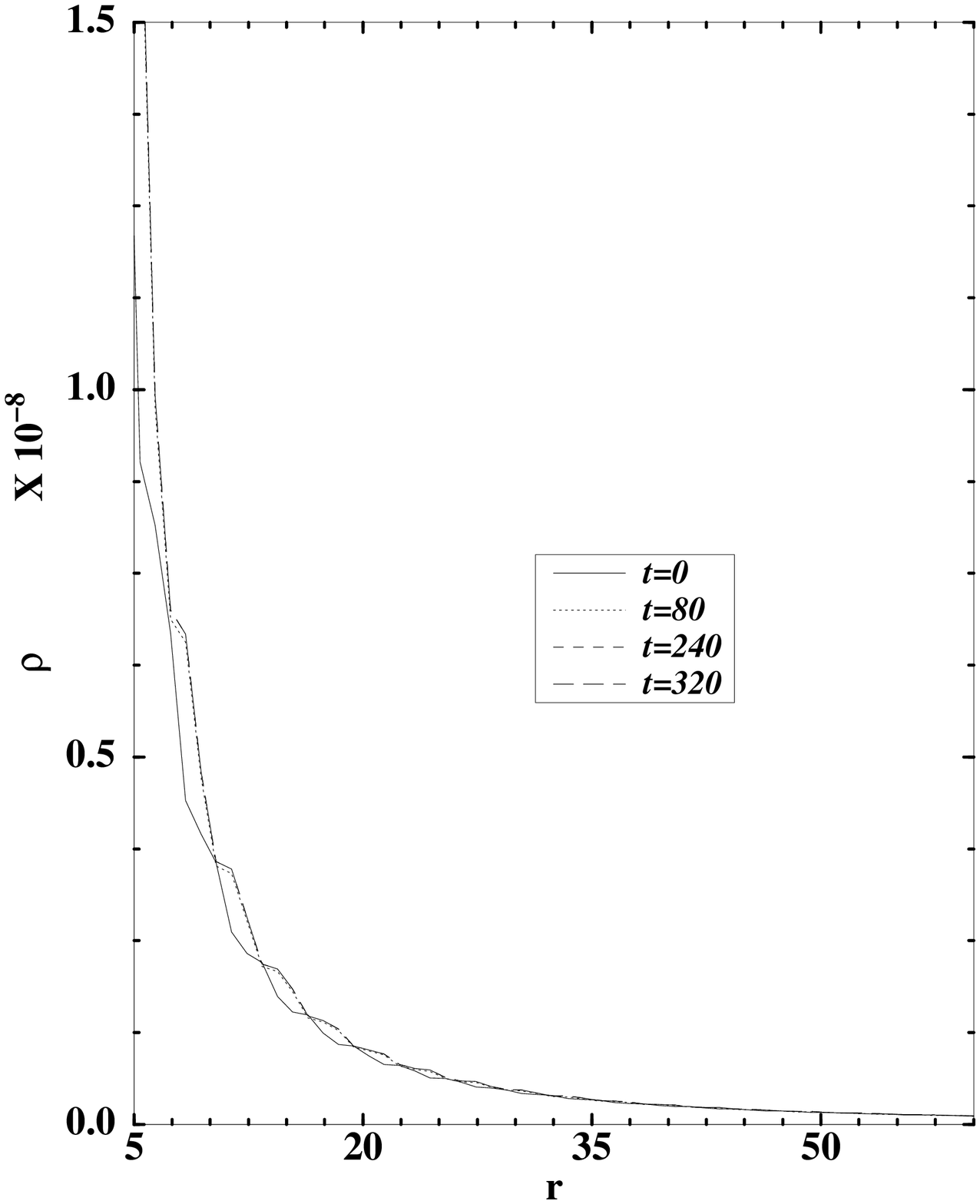}\hskip0.5cm
\leavevmode\epsfysize=7.5cm \epsfbox{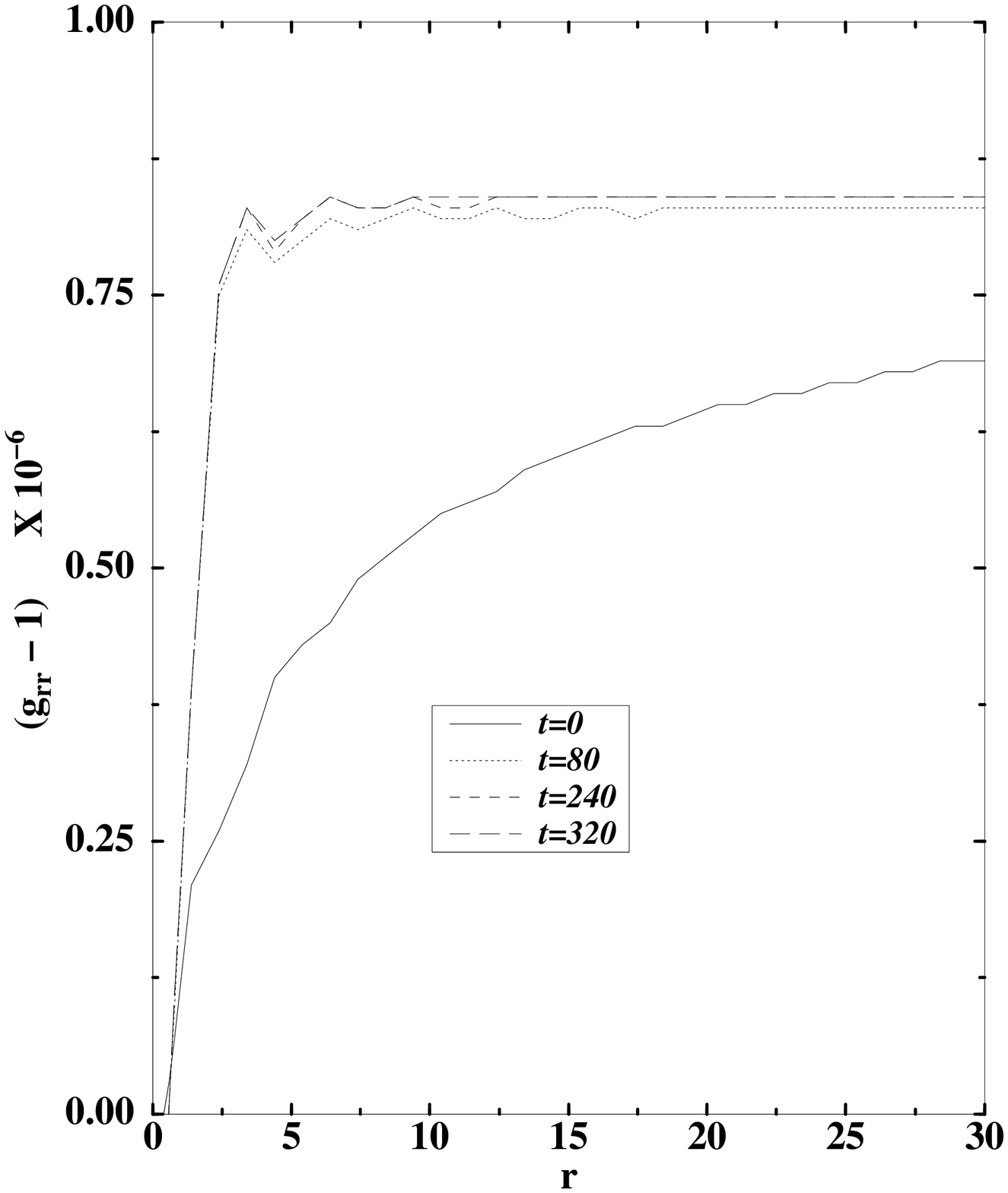}\\
\caption[]
{(a) Left: A Newtonian initial field configuration of the form
$0.001 \cos(r)/(r+1)$ can be regarded in some sense
to be a perturbation of a Newtonian configuration which is proportional to
$\sin(r)/r$. A plot of the density evolution is shown as it settles
to a final stable state. Configurations
of this form with non-Newtonian amplitudes do not settle down.
Since the non-Newtonian configurations do not
have this sinusoidal dependence this is not surprising.
(b) Right: The radial metric evolution for the above configuration
shows its settling in time.}
\end{figure}

\begin{figure}
\centering
\leavevmode\epsfysize=8.5cm \epsfbox{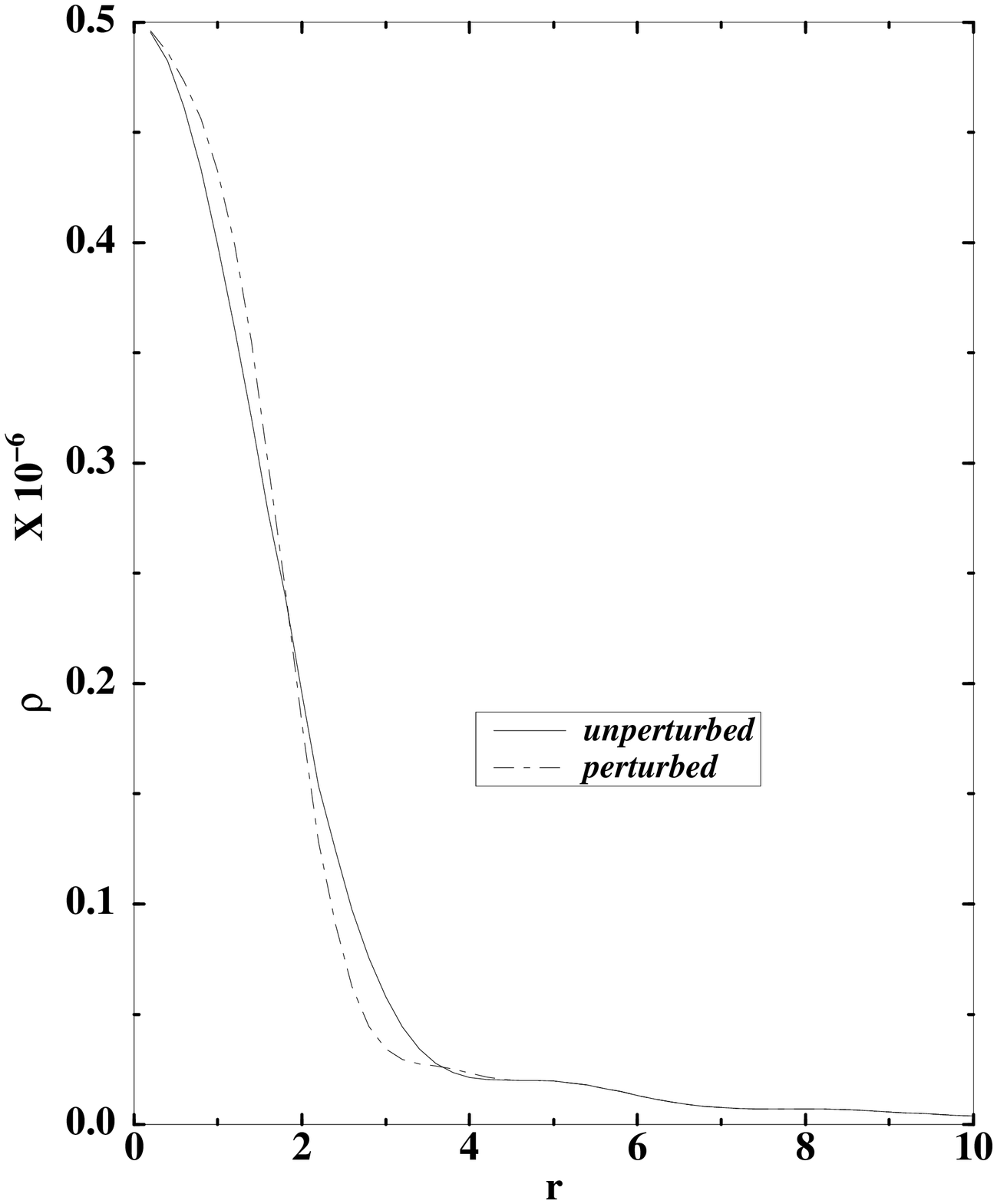}\hskip0.5cm
\leavevmode\epsfysize=8.5cm \epsfbox{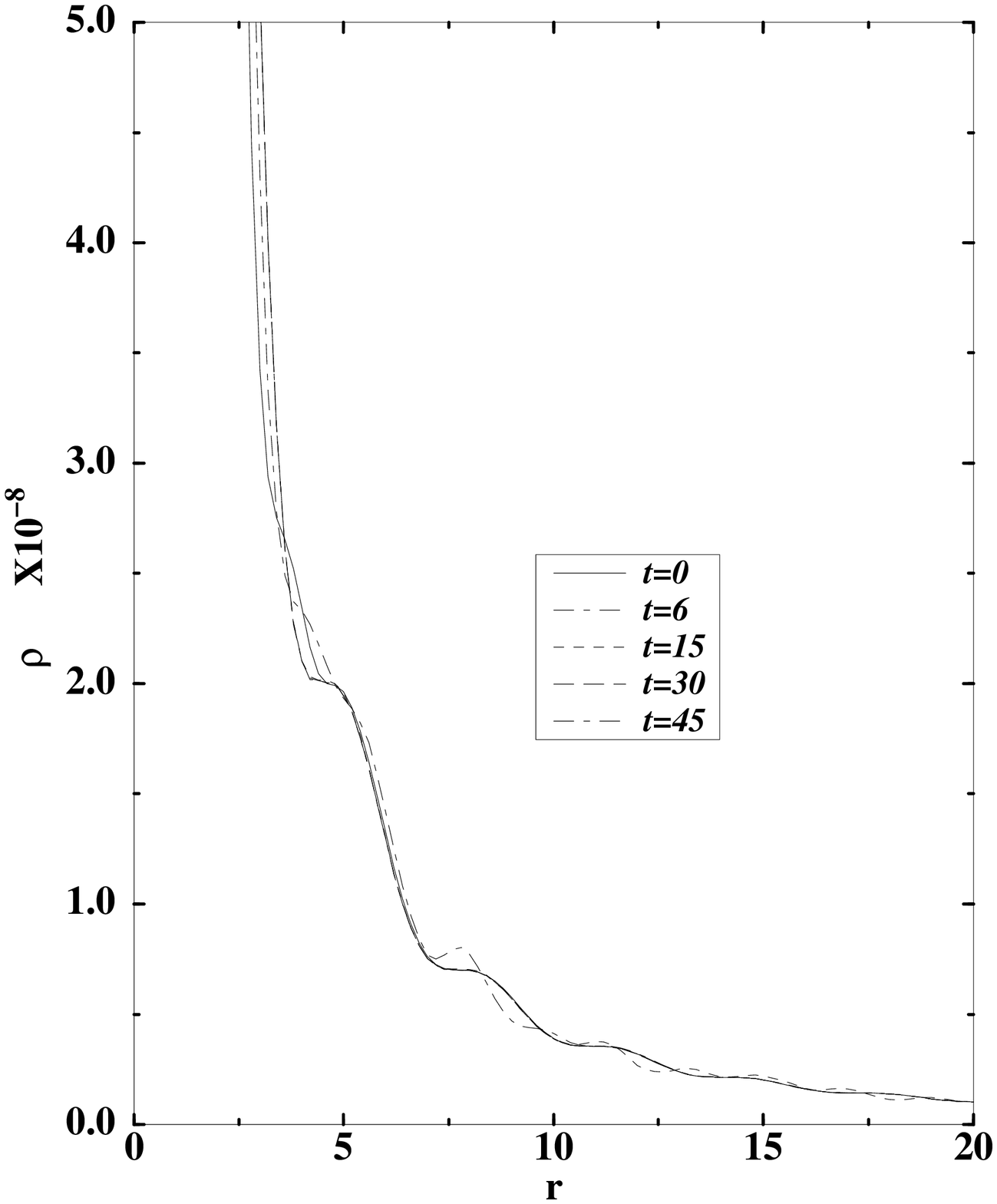}\\
\leavevmode\epsfysize=8.5cm \epsfbox{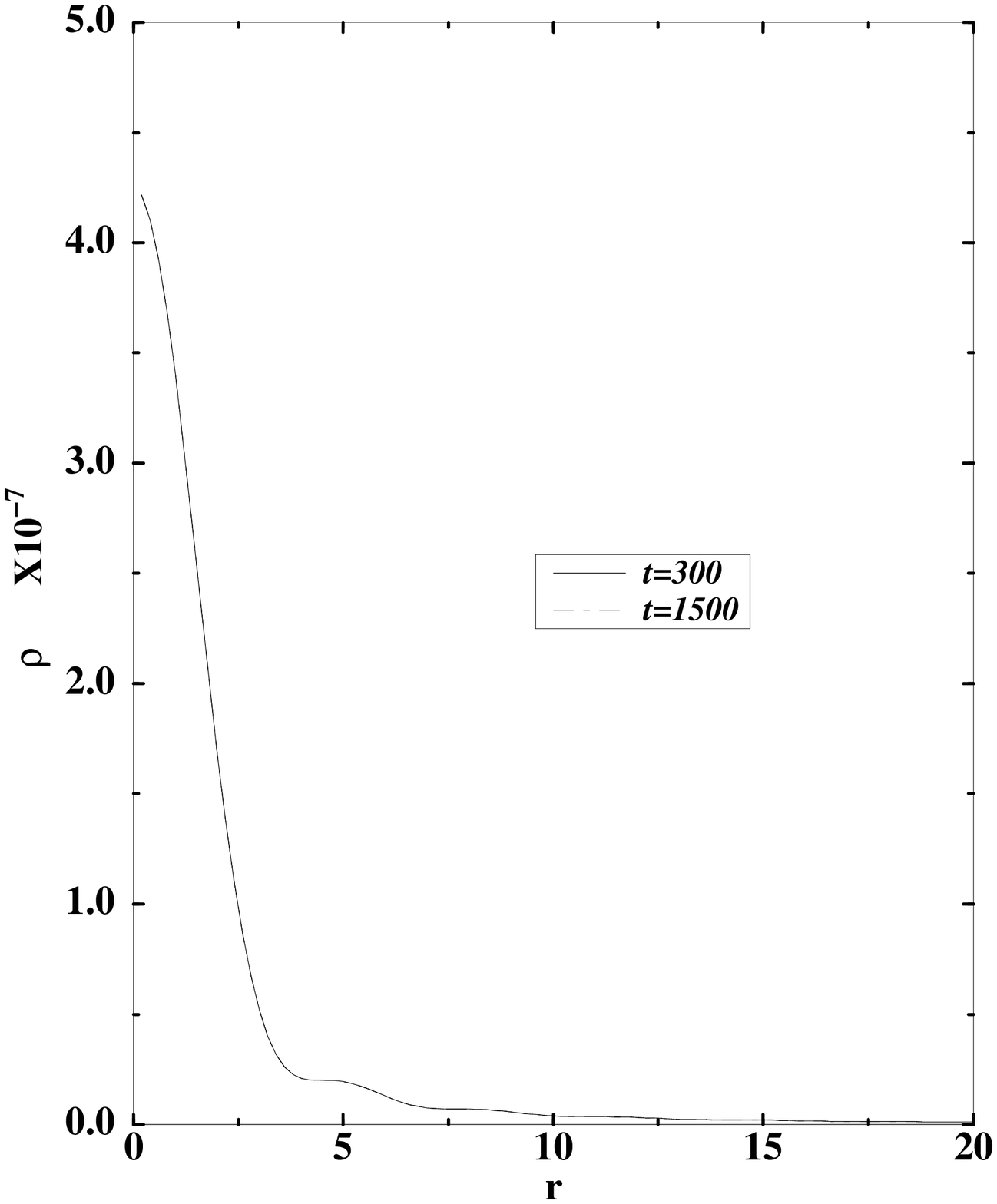}\hskip0.5cm
\leavevmode\epsfysize=8.5cm \epsfbox{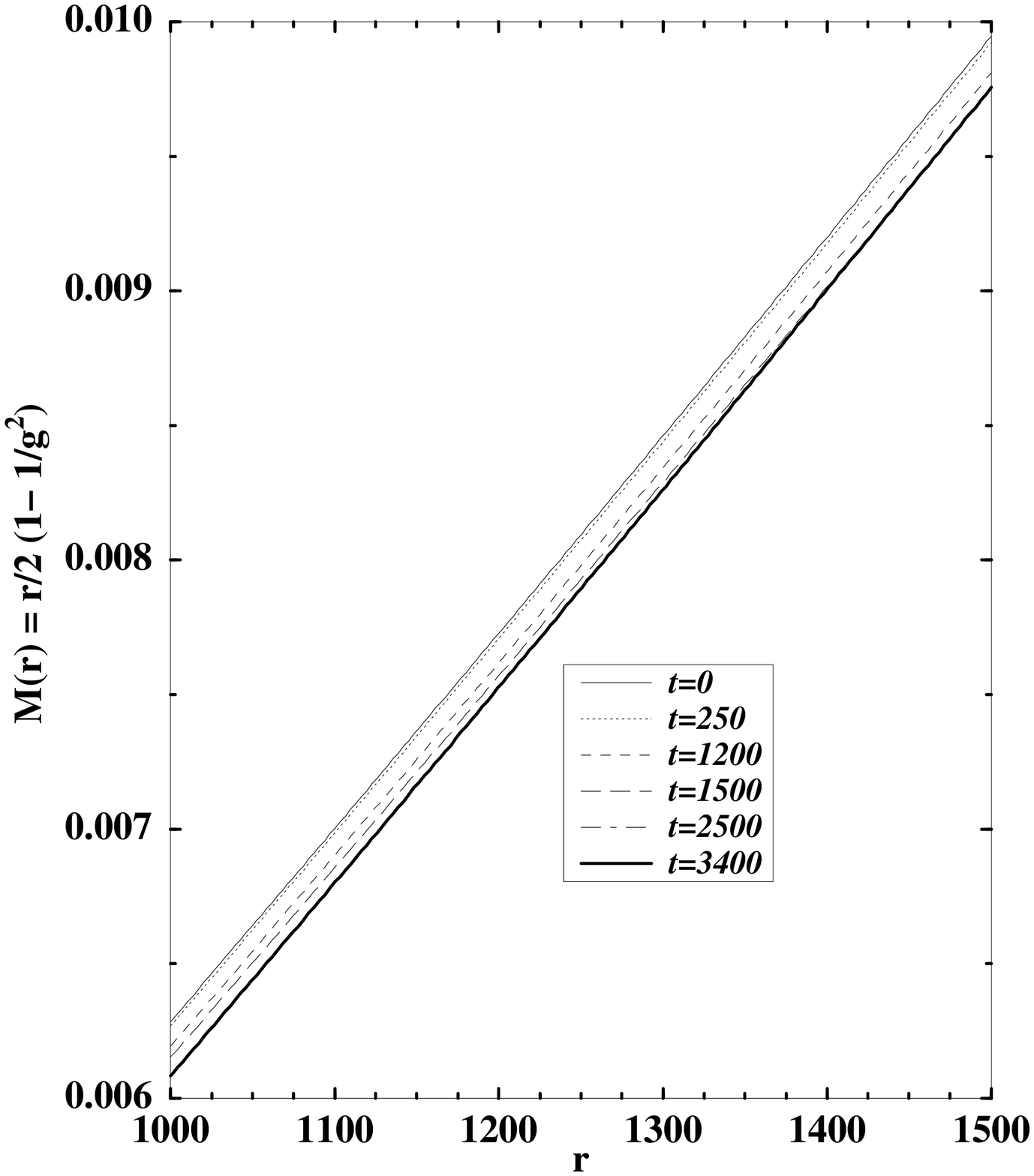}\\
\caption[]
{(a) Above left: An exact Newtonian configuration is perturbed and evolved. 
The initial unperturbed and perturbed densities
are shown. A Gaussian amount of scalar field is removed from the
equilibrium configuration and the
constraint equations are reintegrated to give new metrics before
beginning evolution. 
(b) Above right: The outgoing scalar radiation can be seen in a plot of the
density at early times.
(c) Bottom left: The density profile is shown after it starts to settle down.
Between times of 300 and 1500
very little has changed in the density profile.
(d) Bottom right: The mass is plotted as a function of radius for
different times. In order to enhance the features
only a small part of the radial region is shown. The amount of mass
loss is clearly lessening in time.}
\end{figure}

\begin{figure}
\centering
\leavevmode\epsfysize=7.5cm \epsfbox{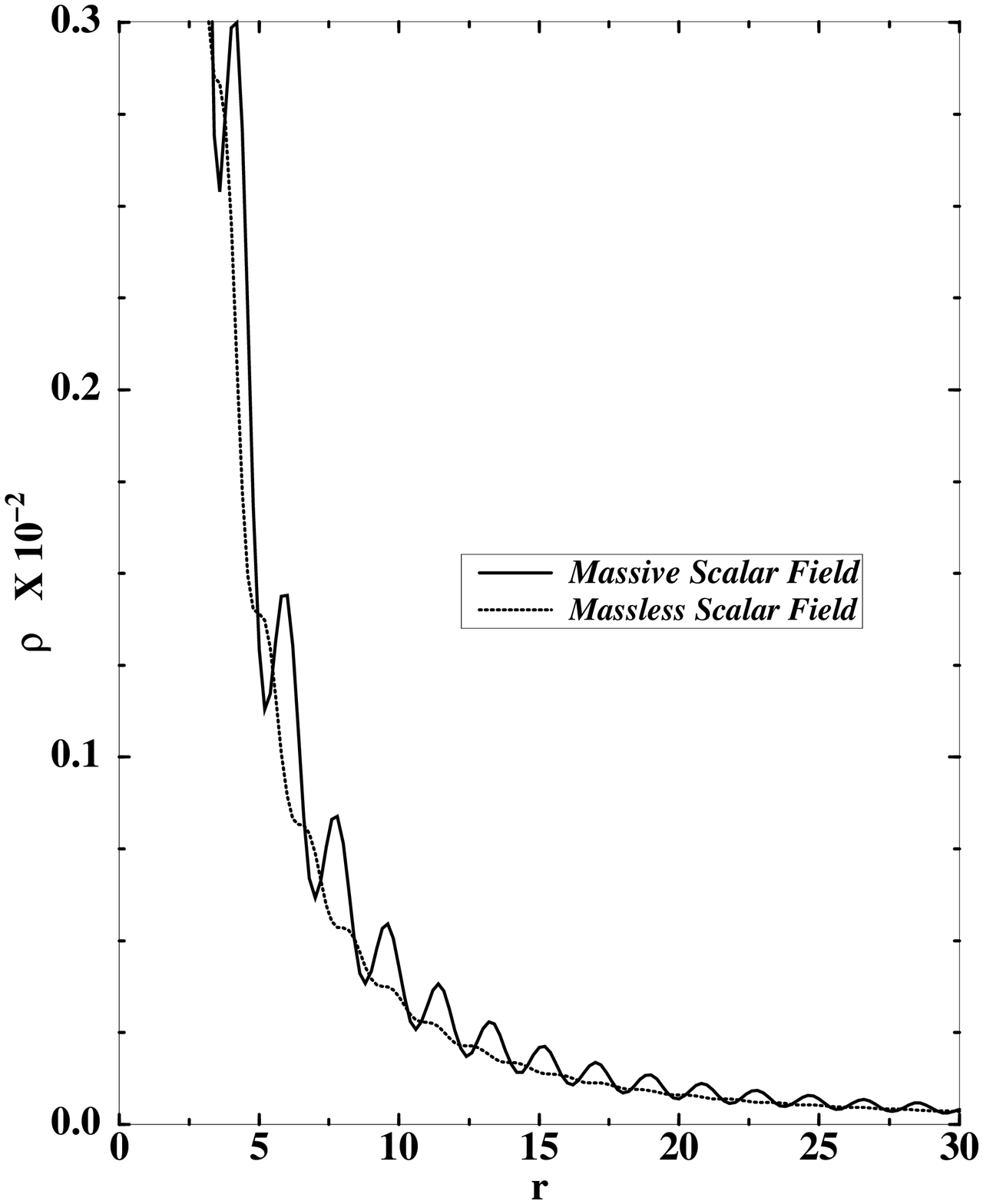}\hskip0.5cm
\leavevmode\epsfysize=7.5cm \epsfbox{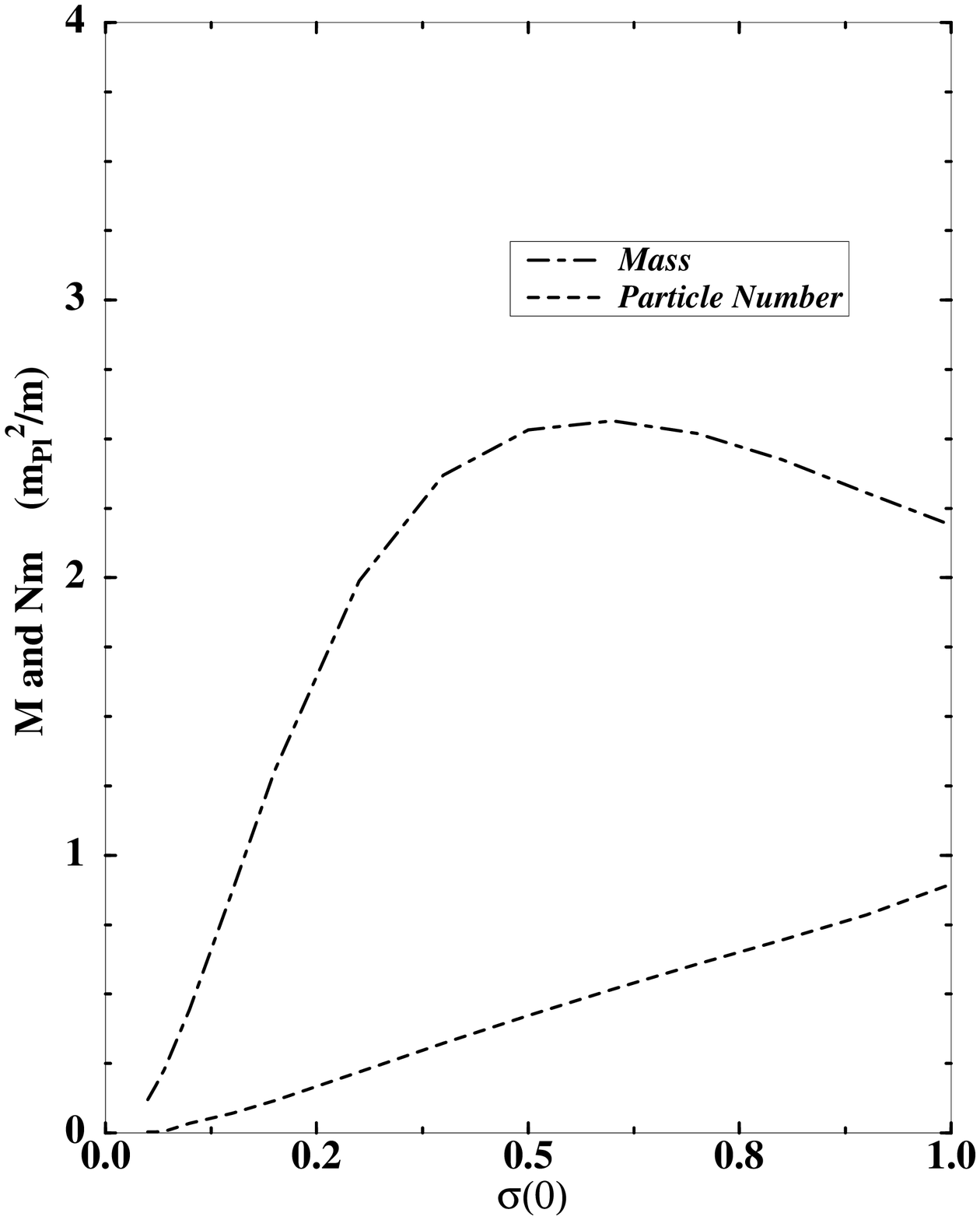}\\
\caption[]
{(a) Left: The density profiles of massive scalar field configurations with
$\omega > m$ is compared to that of
a massless field. The former has maxima and minima while the latter
has saddle points. This
structure may be related to stability issues.
(b) Right: The mass and particle number are plotted against the central density
for boson star oscillators. We find in all cases that the mass is always
greater than the
particle number explaining their instability against a collective dispersion.}
\end{figure}


\begin{table}
\caption[]{Mass values at different radii and different times for the
evolution in Fig.~3. One recognizes that the change in the mass at
fixed radius decreases, hence it proves the settling to a stable
configuration.}
\begin{tabular}{|c|c|c|c|c|}
\multicolumn{5}{c}{Table 1.}   \\
\hline
 & $t=0$ &
$t=500$ & $t=700$& $t=900$
\rule[-0.1in]{0.0in}{0.3in}\\
 \hline
$r=800$ &$0.03589375$&
$0.0350625$ & $0.034625$& $0.034234375 $
\rule[-0.1in]{0.0in}{0.3in}\\
 \hline
$r=1200$ & $0.0575$&
$0.0563125$ & $0.055625$& $0.0550625 $
\rule[-0.1in]{0.0in}{0.3in}\\
 \hline
$r=1600$ & $0.0801875$&
$0.078625$ & $0.0776875$& $0.0769375 $
\rule[-0.1in]{0.0in}{0.3in}\\
\hline
\end{tabular}
\end{table}

\begin{table}
\caption[]{Mass values at different radii and different times for the
evolution in Fig.~5. One recognizes that the change in the mass at
fixed radius decreases, hence it proves the settling to a stable
configuration.}
\begin{tabular}{|c|c|c|c|c|}
\multicolumn{5}{c}{Table 2.}   \\
\hline
 & $t=0$ &
$t=1500$ & $t=2500$& $t=3400 $
\rule[-0.1in]{0.0in}{0.3in}\\
 \hline
$r=500$ &$0.002847$&
$0.00278$ & $0.0026875$& $0.00268375 $
\rule[-0.1in]{0.0in}{0.3in}\\
 \hline
$r=1000$ & $0.00628625$&
$0.00619375$ & $0.00608495$& $0.00608335 $
\rule[-0.1in]{0.0in}{0.3in}\\
 \hline
\end{tabular}
\end{table}

\end{document}